\documentclass[12pt,noshowpacs,nofootinbib,notitlepage,amsmath]{revtex4-2}
\usepackage{setspace}
\allowdisplaybreaks
\usepackage{graphicx,color}
\usepackage[colorlinks=true,citecolor=blue,linkcolor=blue,urlcolor=blue]{hyperref}
\usepackage[charter]{mathdesign}
\usepackage{multirow}
\usepackage{enumerate}
\usepackage{array}
\usepackage{booktabs}

\DeclareSymbolFontAlphabet{\mathcal}{symbols}
\DeclareSymbolFont{symbols}{OMS}{xmdcmsy}{m}{n}
\DeclareSymbolFont{largesymbols}{OMX}{cmex}{m}{n}
\SetSymbolFont{symbols}{bold}{OMS}{xmdcmsy}{b}{n}

\begin{document}  
\title{\color{blue}\Large Ultra-Planckian scattering from a QFT for gravity}

\author{Bob Holdom}
\email{bob.holdom@utoronto.ca}
\affiliation{Department of Physics, University of Toronto, Toronto, Ontario, M5S 1A7, Canada}
\begin{abstract}
Astonishing cancellations take place in the calculation of high-energy scattering cross sections in quantum quadratic gravity, a quantum field theory for gravity. Tree-level differential cross sections that are minimally inclusive behave as $1/E^2$, as desired for a well-behaved UV completion. Such cross sections are calculated for the various spin states of the massless and massive graviparticles. These describe the hard scattering processes that occur in a picture involving the gravitational analog of parton showers. The structure of some of the simpler amplitudes is also explored. Unitarity without positivity is the key property of the perturbative theory.
\end{abstract}

\maketitle 

\section{Introduction}
The action of quantum quadratic gravity (QQG) has $R_{\mu\nu}R^{\mu\nu}$ and $R^2$ terms in addition to the Einstein term. This theory was proposed early as a quantum field theory for gravity when it was found to display some of the desired characteristics of a UV-complete theory; namely, renormalizability and asymptotic freedom \cite{stelle,julve,fradkin}. In addition QQG has an energy bounded from below, in contrast to the classical version of the theory. In this sense QQG could be viewed as an intrinsically quantum mechanical proposal for a fundamental theory. At sub-Planckian energies QQG effectively goes over to general relativity, and GR makes sense both as an effective low-energy QFT and as a classical theory when $\hbar$ is set to zero.

The main obstacle for QQG has been the existence of negative norm states, or ghosts, in the perturbative spectrum. The negative norm for the ghost is synonymous with the wrong overall sign of the ghost propagator, and the perturbation theory becomes populated with odd minus signs. Perhaps most strikingly odd is the existence of negative cross sections. But when the theory is probed a little more and self-energy corrections for the ghost propagator are considered, it is found that the meaning of the apparent negative cross sections changes. In particular the negative norm states are not asymptotic states, and thus the processes that give rise to negative cross sections cannot actually be physically realized. We shall return to the origin of negative cross sections and their meaning in Sec.~\ref{s3}.

The path integral definition of QQG is standard in the sense that the $i\varepsilon$ prescription for propagators is the Feynman prescription. This leads to a renormalizable perturbation theory where ghosts propagate with positive energy. At any finite order in perturbation theory the loop diagrams have standard analytic structure. The imaginary part of the ghost self-energy, due to normal light particles going on-shell, has the standard sign. But the resummation to produce a dressed ghost propagator produces an unusual result. Because of the overall minus sign of the ghost propagator, the result is an imaginary term in the inverse propagator that is opposite in sign to the $i\varepsilon$. This changes the propagation of the ghost in a fundamental way. Not only does the ghost decay, but it decays backward in time. In turn this means that the ghost cannot be considered an asymptotic state in any sense. $S$-matrix elements can be calculated in the perturbative theory, but those involving ghosts do not correspond to the scattering of asymptotic states. For a thorough discussion of backward in time decay see \cite{Donoghue:2019ecz,Donoghue:2020mdd}.

At this point one is tempted to more directly take into account the true nature of ghost propagation within perturbation theory, which must then be reformulated in terms of the dressed propagator. This type of approach has a long history, going back to earlier work by Lee and Wick on higher-derivative theories \cite{leewick} and continuing in a more modern context \cite{Grinstein:2008bg}. Recently unitarity was proved in this type of framework, where cutting rules are adjusted to account for the use of the dressed ghost propagator and imaginary parts of diagrams are related to cuts involving only standard stable particles \cite{Donoghue:2019fcb}. An approach to high-energy scattering with the dressed ghost propagator appearing as an internal line of tree graphs was considered in \cite{Salvio:2018kwh}.\footnote{There are also efforts to rewrite quantum mechanical models with negative norm states in such a way that the negative probabilities do not appear \cite{Bender:2007wu,Salvio:2018crh,Strumia:2017dvt,Salvio:2020axm}, but given that the ghost in the QFT is not an asymptotic state, we feel that this is not necessary.}

We shall instead continue along the path of the original perturbative QQG with the corresponding bare propagators. As we have mentioned, the perturbative expansion will have standard analytic structure with no need of unusual shifts of contours. While we have to keep in mind that the ghosts are not asymptotic states, they are among the fundamental degrees of freedom in which the theory is expressed. There is an important application where it is appropriate to use the fundamental perturbative description, and this is in the treatment of high-energy scattering in an asymptotically-free theory. The high-energy scattering of protons at the LHC for instance is largely described in terms of the interactions of quarks and gluons, the fundamental degrees of freedom of perturbative QCD; like the ghosts, quarks and gluons are also not asymptotic states.

On the other hand quarks and gluons are not ghosts. Because ghosts are present among the perturbative degrees of freedom of QQG, it is often said that QQG suffers from a lack of unitarity. This is not quite right, and a more precise statement is that it lacks positivity. Unitarity is still present, and the lack of positivity is the single foundational property that the theory lacks. The backward in time decay for instance is a consequence. The combination of the positive graviton propagator and the negative ghost propagator leads to $(q^2)^{-2}$ behavior at large $|q^2|$, and so the lack of positivity is intrinsic to the renormalizability of the theory. It also shows up in the way the $(q^2)^{-2}$ behavior is not compatible with the Kallen-Lehmann representation of a propagator in terms of a positive definite spectral function. Unitarity without positivity and the resulting optical theorem is discussed in Sec.~\ref{s3}. We shall argue in this paper that the lack of positivity does not pose a problem for the description of ultra-Planckian scattering. On the contrary, it is necessary to produce good high-energy behavior.

\section{Preview of high-energy behavior}

High-energy scattering in QQG, at energies well above Planck energies, appears to be poorly behaved. This can be seen by simply considering the tree-level vertices arising from the quadratic-in-curvature part of the action. (We display this action in Sec.~\ref{s5}.) There are vertices quartic in the gravitational field that involve four derivatives. This leads to contributions to amplitudes with an $E^4$ behavior, before cancellations between diagrams are considered. In fact, it seems even worse that this. We are interested in scattering among the various polarization states of the gravitational degrees of freedom, the graviparticles. These are the two spin-2 polarizations of the massless graviton, the two spin-2, two spin-1 and one spin-0 polarizations of the massive ghost, and the spin-0 polarization of the massive graviscalar. In constructing the amplitudes, we need the polarization tensors of these various states (see Appendix \ref{a1}). The spin-1 and spin-0 polarization tensors involve one and two factors of a longitudinal vector $l_\mu$ and thus one or two factors of $E/m$. Thus an amplitude involving four spin-0 graviparticles for example, would seem to have an additional factor of $(E/m)^8$.

Now let us describe our results a little, just to show how wrong these expectations are. For any 2-to-2 process there are four diagrams to add, three exchange-type diagrams with the exchange in the $s$, $t$, and $u$ channels, and one contact-type diagram that we have already mentioned. From such an amplitude we can obtain the exclusive differential cross section for the scattering of some particular choice of polarizations states. Let us first consider scattering involving only spin-1 and/or spin-0 states. What we find is that any such differential cross section falls like $1/E^2$, and in some cases even faster. What this means is that astonishing cancellations are occurring between the four diagrams. We shall refer to these cancellations as intrinsic cancellations, to differentiate them from another type of cancellation that involves the spin-2 states.

When scattering involves the spin-2 states, the intrinsic cancellations can still leave $E^2$ or $E^0$ behavior in exclusive differential cross sections. So let us consider differential cross sections that are slightly more inclusive. That is, for every spin-2 external state with whatever polarization, we sum over two exclusive differential cross sections, one where the spin-2 state is the graviton and the other where it is the spin-2 component of the ghost. Thus for $n$ spin-2 external lines, we sum $2^n$ exclusive differential cross sections. When this is done then further astonishing cancellations occur, and the resulting minimally-inclusive differential cross section again falls like $1/E^2$, and in some cases even faster. Since it is only the external spin-2 masses that are changing in the sum, along with the changing signs, we refer to these cancellations as mass cancellations.

These intrinsic and mass cancellations appear to be a nontrivial consequence of the unitarity and the renormalizability of the theory. We end up with a whole set of differential cross sections that are bounded at high energies by $1/E^2$ behavior, just as for other sensible UV-complete quantum field theories. These exclusive or minimally-inclusive differential cross sections will typically turn out to be positive, but not always. If not, then they do become positive when the inclusiveness is increased further. This then folds into a gravitational parton shower picture for ultra-Planckian scattering, where it is the inclusive cross sections that are needed. 

We begin in Sec.~\ref{s3} with a brief discussion of unitarity in QFT in the presence of negative norm states. In Sec.~\ref{s5} we define QQG and what it is that we are calculating and in Sec.~\ref{s6} we give the differential cross section results. In Sec.~\ref{s7} we describe some of the simpler amplitudes. Then after the discussion of gravitational parton showers in Sec.~\ref{s4} we conclude in Sec.~\ref{s8}.

\section{Unitarity without positivity}\label{s3}

In quantum field theory, a scattering or decay of an initial state $|i\rangle$ to a final state $|j\rangle$ has a differential probability $dP$ of the form
\begin{align}
dP&= d\Pi_f\frac{|\langle f|S|i\rangle|^2}{\langle f|f\rangle\langle i|i\rangle},\label{e1}\\
\langle X|X\rangle=&\prod_{j\in X}\eta_j\, 2 E_j, \quad  \quad d\Pi_X=\prod_{j\in X}\frac{d^3p_j}{(2\pi)^3}.\nonumber
\end{align}
Space and time volume factors $V$ and $T$ are omitted since they cancel when going from the differential probability $dP$ to a differential cross section $d\sigma$ or decay rate $d\Gamma$. The only thing new is $\eta_j=\pm1$, which is negative for a ghost. When the total number of ghosts in the initial and final states is odd, the product of norms in the denominator (\ref{e1}) is negative and the resulting cross section or decay rate is negative.

The first thing we learn from (\ref{e1}) is that a ghost has a negative decay rate, given that its Planck-scale mass gives many possible decays to normal particles. A negative decay rate corresponds to the abnormal sign of the imaginary part of self-energy correction in the dressed propagator, and it is such a propagator that implies that the ghost decays backward in time \cite{Grinstein:2008bg}\cite{Holdom:2019ouz}\cite{Donoghue:2019ecz,Donoghue:2020mdd}. Thus this behavior is a consequence of the lack of positivity. Ghosts propagate positive energy backwards in time before decaying to normal particles that then appear in the final state. Particles that decay backward in time cannot be considered asymptotic states even in some approximate sense, as would be the case for a slightly unstable normal particle. In fact ghosts may be maximally unstable, so that the time scale of the acausality could be as short as the Planck time.\footnote{QQG has the Einstein term in the action, but this could be viewed as modeling the dynamical generation of this term through dimensional transmutation. That is, the Planck mass and the Einstein term could be generated when the asymptotically-free coupling grows strong \cite{Holdom:2015kbf,Holdom:2016xfn}. In this case the masses and lifetimes of the massive graviparticles are set by the Planck scale.}

Unitarity can coexist with negative probabilities. Unitarity is defined by $S S^\dagger\equiv S\mathbb{1}S^\dagger=\mathbb{1}$ along with a representation of the identity (the completeness relation) that depends on the norms,
\begin{align}
\mathbb{1}=\sum_X\int d\Pi_X \frac{|X\rangle \langle X|}{\langle X|X \rangle}.
\end{align}
The standard definitions $S=\mathbb{1}+i{\cal T}$ and $\langle f|{\cal T}|i\rangle = (2\pi)^4\delta(p_i-p_f)\langle f|{\cal M}|i\rangle$ then lead to the optical theorem,
\begin{align}
\frac{2{\rm Im}\langle i|{\cal M}|i\rangle}{\langle i|i\rangle}=\sum_X\int d\Pi_X (2\pi)^4 \delta^4(p_i-p_X)\frac{|\langle X|{\cal M}|i\rangle|^2}{\langle X|X\rangle\langle i|i\rangle}.\label{e17}
\end{align}
Both sides have been divided by $\langle i|i\rangle$, since then the various terms on the right can be identified with probabilities, as in (\ref{e1}). When the total number of ghosts in the state $i$ and a particular state $|X\rangle$ is odd, then that term in the sum over states $X$ is negative. It is clear that these signs are part of the description of unitarity and are not a problem with unitarity. What has been lost is the positivity of the individual terms in the sum, or in other words, there are both positive and negative probabilities. Unitarity without positivity could perhaps be called something else, such as pseudounitarity or generalized unitarity, but we don't see the need for that.

The optical theorem (\ref{e17}) shows the link between a negative sign propagator appearing in the calculation of the amplitude $\langle i|{\cal M}|i\rangle$ and a negative norm $\langle X|X\rangle$ appearing on the rhs. Partial cancellations are occurring on both sides. It is these types of cancellations that will play a role in controlling high-energy behavior. Feynman foresaw the utility of this type of situation in \cite{Feyn} where he argued that negative probabilities occurring in intermediate stages of calculation are okay, as long as the individual negative probabilities cannot be physically realized.\footnote{He concludes his paper by writing ``It may have applications to help in the study of the consequences of a theory of this kind by Lee and Wick.''} 

We claim that it is the inclusive cross sections that are of interest for a gravitational parton shower picture for ultra-Planckian scattering. We borrow the term `parton' from the parton shower picture that emerges from the QCD description of high-energy scattering. Here quarks and gluons are the relevant degrees of freedom even though they do not exist on-shell, i.e.~they are not the physical asymptotic states. The partonic cross sections must be suitably inclusive to be dual to the cross sections for the physical states. We continue this discussion in section \ref{s4}.

\section{Partially inclusive differential cross sections}\label{s5}

We are considering high-energy scattering among graviparticles, which include the massless graviton $g$, the ghost $G$ of mass $m_G$, the graviscalar $S$ of mass $m_S$. The action of QQG may be expressed in terms of the masses,
\begin{align}
S&=-\frac{1}{16\pi G}\int d^4x \sqrt{-g}\left(R+ \frac{R_{\mu\nu}R^{\mu\nu}-\frac{1}{3}R^2}{m_G^2}-\frac{R^2}{6m_S^2} \right).
\label{e19}\end{align}
These masses are naturally of order the Planck mass. The couplings $Gm_G^2$ and $Gm_S^2$ run logarithmically due to renormalization effects, but this is not relevant for our tree-level calculations.\footnote{The graviscalar $S$ is associated with the $R^2$ term and a coupling that is not asymptotically-free. In \cite{Salvio:2017qkx} it is argued that the theory flows to a pure conformal theory in the UV where $S$ decouples from the rest of the theory. In our tree-level calculations we choose to carry $S$ along with an independent mass $m_S$.} We shall also include a real scalar $\phi$ of mass $m$ with minimal coupling to gravity. Scattering involving two $\phi$'s and two graviparticles will turn out to be closely related to some of the four graviparticle results.

The graviparticle propagator is
\begin{align}
G_{\mu\nu\rho\sigma}&=i16\pi G\left(-\frac{2m_G^2}{q^2(q^2-m_G^2)}P^2_{\mu\nu\rho\sigma}+\frac{m_S^2}{q^2(q^2-m_S^2)}P^0_{\mu\nu\rho\sigma}\right),\label{e11}\\\nonumber
&P^2_{\mu\nu\rho\sigma}=\frac{1}{2}(\theta_{\mu\rho}\theta_{\nu\sigma}+\theta_{\mu\sigma}\theta_{\nu\rho})-\frac{1}{3}\theta_{\mu\nu}\theta_{\rho\sigma},\\\nonumber
&P^0_{\mu\nu\rho\sigma}=\frac{1}{3}\theta_{\mu\nu}\theta_{\rho\sigma},
\quad\theta_{\mu\nu}=\eta_{\mu\nu}-\frac{q_\mu q_\nu}{q^2}.
\end{align}
The partial fraction decomposition of the terms in (\ref{e11}) yields the graviton, ghost, and graviscalar poles and their relative signs. Possible gauge-dependent additions to this propagator do not contribute to our tree-level calculations of on-shell amplitudes. Our calculations are done for individual amplitudes such that the polarization tensors are contracted from the beginning. These polarization tensors are given in Appendix \ref{a1}.

We shall focus on various partially-inclusive differential cross sections. We work in the center-of-momentum (CoM) frame and for the process $AB\to CD$ we use the variables $E\equiv\sqrt{s}/2=(E_A+E_B)/2=(E_C+E_D)/2$ and $\theta$, the scattering angle between $A$ and $C$. Some partial sum over processes takes the form
\begin{align}
\sum_{\rm (pol,mass)}^{\rm partial}(-1)^{n_G} \frac{p_f(E)}{p_i(E)}|{\cal M^{\rm (pol,mass)}}(E,\theta)|^2
\label{e10}.\end{align}
${\cal M}^{\rm (pol,mass)}$ denotes an on-shell amplitude for a particular choice of spin, polarization and mass for each external graviparticle. $n_G$ is the number of ghosts among $A,B,C,D$. The ratio of 3-momentum magnitudes arises from the phase-space/flux factor for a differential cross section. The dependence on the masses of $A$, $B$, $C$, and $D$ for each process in (\ref{e10}) is implicit.

We are interested in the high-energy behavior and we expand the amplitudes in powers of $1/E$ before squaring. The leading term in $1/E$ of (\ref{e10}) is proportional to the leading term of some dimensionless differential cross section, $s d\sigma/d\Omega$, in the CoM frame. We shall refer to this leading term as a high-energy partially-inclusive dimensionless differential (\textbf{hepidd}) cross section, which we write as
\begin{align}
\left.E^2\frac{d\sigma^{\rm na}}{d\Omega}(E,\theta)\right|_{E\to\infty}
.\end{align}
$\sigma^{\rm na}$ denotes a cross section that is ``not averaged''; that is, it has not been divided by the number of different initial states in the partial sum. This is convenient for the continued combining of hepidd cross sections; initial state averaging can be done for some final inclusive cross section of interest.

For any particular choice of the set of spin/polarizations, the mass cancellations only require a summation over the two choices of mass for each spin-2 particle. But we shall consider more inclusive sets of processes simply to reduce the number of sets to consider. If $C$ and $D$ are distinct graviparticles then we include the interchange of $C$ and $D$, and the same for $A$ and $B$. These interchanges give terms in (\ref{e10}) that are related by $\theta\to\pi-\theta$, and so their sum gives a result that is $\theta\leftrightarrow\pi-\theta$ symmetric. In other words our hepidd cross sections are $t\leftrightarrow u$ symmetric.\footnote{Total cross sections are obtained by integrating over half the hemisphere to avoid double counting.} We shall also include the reverse processes, for example $CD\to AB$, if they are distinct. We just need to reverse the ratio, $p_f/p_i\to p_i/p_f$, for each reversed process. Reversing this ratio changes the $1/E$ expansion in such a way that the mass cancellations still occur while possibly giving a different contribution to the hepidd cross section.

The sets of processes we consider can then be labeled by their spin content $(ab,cd)$ where each of $a,b,c,d$ can be any of the spin components of the ghost, 2, 1, 0, or it can be the graviscalar $S$. The $ab$ and $cd$ spins can appear in the initial and final states via all the possible interchanges as we have described. We must also consider all possible placements of the $e$ and $o$ polarization labels on the nonzero spins (see Appendix \ref{a1}), as well the two masses for each spin-2. Each such choice gives one amplitude and one term in (\ref{e10}). The calculation of each such amplitude involves expanding some very large expression to high order in $1/E$ to find the leading nonzero orders. This is where the intrinsic cancellations are occurring.

All contributions to hepidd cross sections not involving spin-2 particles have $E^0$ behavior due to intrinsic cancellations (and thus $1/E^2$ behavior for dimensionful cross sections). Hepidd cross sections involving spin-2 particles (as many as four) have contributions with at most an $E^4$ behavior. For an $E^4$ behavior there are at least two spin-2 particles, and the resulting sum over the different mass combinations yield cancellations that bring the behavior down to $E^0$. The leading, next-to-leading and next-to-next-to-leading terms in the amplitudes are needed to find the $E^0$ result. In other cases an $E^2$ behavior is brought down to $E^0$. The different terms being added to find these mass cancellations typically have very different $\theta$ dependence. These cancellations were the most surprising to us.

For the processes labeled by $(22,22)$, one of the $E^4$ contributions is graviton-graviton scattering, $gg\to gg$. It was noticed in \cite{Dona:2015tra} that that QQG and GR give the same result for $gg\to gg$. In fact it was seen that this is true of any tree-level process that does not have the massive states of QQG as external particles. As a check we use QQG to reproduce the GR cross sections as given in \cite{Berends:1974gk} for $gg\to gg$ [see (\ref{e3})] and $g\phi\to g\phi$. We have also considered the effect of removing the Einstein term from the QQG action. The remaining quadratic terms are not pure Weyl, but nevertheless the $gg\to gg$ amplitude vanishes in this case while the $g\phi\to g\phi$ amplitude remains the same.

\section{Gravi particle scattering results}\label{s6}

\def\arraystretch{.6}
\newcolumntype{x}[1]{>{\vfil$\displaystyle} p{#1} <{$\vfil}} 

\begin{table}[h]
\centering
\begin{tabular}{x{1.7cm}|x{12cm} }
\midrule
   & \frac{E^2}{G^2m_G^4}\left.\frac{d\sigma^{\rm na}}{d\Omega}(E,\theta)\right|_{E\to\infty}\mbox{ with }\; {\Large\substack{C=\cos(\theta)\\S=\sin(\theta)}},\; x=\frac{m_S}{m_G}  \\ \midrule\midrule
  (02,02)^\dagger & -2\frac{C^{8}-56 C^{6}-150 C^{4}-48 C^{2}-3}{S^{8}}\\ \midrule
  (01,01)^\dagger & 4\frac{ \left(C^{2}+3\right)^{2} \left(C^{4}+6 C^{2}+1\right)}{S^{8}}\\ \midrule
  (02,01)^\dagger & -32\frac{ \left(3 C^{2}+1\right)}{S^{4}}\\ \midrule
  (00,22)^\dagger & -\frac{S^{2} \left(15 C^{2}-7\right)}{16}\\ \midrule
  (00,11)^\dagger & \frac{\left(5 C^{2}-1\right)^{2}}{16}\\ \midrule
  (00,21)^\dagger & 3 C^{2} S^{2} \\ \midrule
  (SS,00) & \frac{\left(3 C^{2}+4 x^{2}-1\right)^{2}}{72}\\ \midrule
  (S0,S0) &  2 \left(\frac{x^{2}}{3}+\frac{1}{6}+\frac{2 C +2}{\left(C -1\right)^{2}}\right)^{2}+(C\to -C)\\ \midrule
  \end{tabular}
\caption{The high-energy partially-inclusive dimensionless differential cross sections for processes that have some relation to two $\phi$ processes. Notice that we have divided by $G^2m_G^4$. $^\dagger$In these cases the corresponding result for $0\to S$ can be obtained by multiplying by four.}\label{t2}
\end{table}

\begin{table}[h]
\centering
  \begin{tabular}{x{1.7cm}|x{12cm} }
\midrule
   & \frac{E^2}{G^2m_G^4}\left.\frac{d\sigma^{\rm na}}{d\Omega}(E,\theta)\right|_{E\to\infty}\mbox{ with }\; {\Large\substack{C=\cos(\theta)\\S=\sin(\theta)}},\; x=\frac{m_S}{m_G}  \\ \midrule\midrule
  (22,22) & \frac{5 C^{12}+140 C^{10}+349 C^{8}+240 C^{6}-193 C^{4}+100 C^{2}+383}{S^{8}}    \\ \midrule
  (21,21) & \frac{64 C^{12}+403 C^{10}-653 C^{8}+1350 C^{6}-66 C^{4}+391 C^{2}+559}{2 S^{8}}  \\ \midrule
  (11,11) & \frac{81 C^{12}+234 C^{10}-425 C^{8}+1708 C^{6}-753 C^{4}+2570 C^{2}+681}{16 S^{8}} \\ \midrule
  (22,21) & -4\frac{6 C^{8}+99 C^{6}+87 C^{4}+33 C^{2}-97}{S^{4}} \\ \midrule
  (21,11) & -2\frac{ 11 C^{8}+68 C^{6}-30 C^{4}+68 C^{2}+11}{S^{4}} \\ \midrule
  (22,11) & -5\frac{19 C^{6}+103 C^{4}-47 C^{2}-11}{8 S^{2}} \\ \midrule
  (00,00) & \frac{\left(C^{6}-C^{4}+47 C^{2}+17\right)^{2}}{64 S^{8}} \\ \midrule
  (SS,SS) & \frac{\left(C^{6}-8 x^{2} C^{4}-C^{4}+16 x^{2} C^{2}+47 C^{2}-8 x^{2}+17\right)^{2}}{4 S^{8}} \\ \midrule
  (SS,S0) & -8 x^{4}\\ \midrule
  \end{tabular}
\caption{The hepidd cross sections for the remaining processes. Processes that fall even faster with $E$ are not listed.}\label{t3}
\end{table}

First we look at processes involving two and only two $S$'s. We find a surprising relation to processes where each $S$ is replaced by the real scalar $\phi$. The amplitudes for $SS\to AB$ and $AS\to BS$ for example are a factor of two times the amplitudes for $\phi\phi\to AB$ and $A\phi\to B\phi$ respectively, when the $\phi$ mass $m$ is set equal to the $S$ mass $m_S$. We say this is surprising only because the calculations are so different, with the $S$ amplitude calculation being significantly more complex. The fact that we are finding these relations is apparently a good check on our calculations. The $\phi$ amplitudes are actually sufficient to construct all the amplitudes for the processes in the sets labeled by $(SS,ab)$ and $(Sa,Sb)$. Note that the hepidd cross sections for the $(Sa,Sb)$ processes have a $t\leftrightarrow u$ symmetry, while the hepidd cross sections for the $A\phi\to B\phi$ processes do not. We give the hepidd cross sections for the processes with two $\phi$'s in Appendix \ref{a2}.

Next we consider processes involving two and only two $0$'s (two spin-0 components of the ghost), and no $S$'s. Here we find that the hepidd cross sections are $1/4$ those with each $0$ replaced by $S$. A correspondingly simple statement about the individual amplitudes does not necessarily hold in this case. We note also that while processes involving $S$ can be expected to show a dependence on $m_S$ as well as $m_G$, these hepidd cross sections showing the simple relation under $0\leftrightarrow S$ have no dependence on $m_S$.

We list the hepidd cross sections that have these relations between $0$, $S$, and $\phi$ in Table \ref{t2}. The actual dimensionful-differential cross sections are $G^2m_G^4/E^2$ times the entries in the table. The remaining processes are genuinely new and the results for these processes are shown in Table \ref{t3}. Notice that the last entry in Table \ref{t3} for $(SS,S0)$ is the only one with an odd number of $S$'s or $0$'s. Other such processes, such as $(22,20)$, $(22,10)$, $(21,10)$, $(11,10)$, $(20,00)$, $(10,00)$, $(2S,SS)$, $(1S,SS)$, give hepidd cross sections that fall off even faster, that is like $E^{-2}$ rather than $E^0$, and so we do not include them. Amplitudes involving a single $S$ identically vanish. Amplitudes with three gravitons and a massive graviparticle also identically vanish.

\def\arraystretch{1.3}
\begin{table}[ht]
\centering
\begin{tabular}{|c|c|c|c|c|}
\hline
process label & power of $E$ & power of $\sin(\theta)$ & residue of $\theta^{-8}$ & value at $\theta=\pi/2$ \\\hline\hline
(22,22) & 4 & $-8$ & $2^{10}$ & 383 \\\hline
(21,21) & 4 & $-8$ & $2^{10}$ & 559/2 \\\hline
(11,11) & 0 & $-8$ & $2^8$ & 681/16 \\\hline
(22,21) & 2 & $-4$ & $-$ & 388 \\\hline
(21,11) & 2 & $-4$ & $-$ & $-22$ \\\hline
(22,11) & 4 & 0 & $-$ & 55/8 \\\hline
(02,02)$^\dagger$ & 4 & $-8$ & $2^{9}$ & $6$ \\\hline
(01,01)$^\dagger$ & 0 & $-8$ & $2^{8}$ & 18 \\\hline
(02,01)$^\dagger$ & 2 & $-4$ & $-$ & $-32$ \\\hline
(00,22)$^\dagger$ & 4 & $0$ & $-$ & 7/16 \\\hline
(00,21)$^\dagger$ & 2 & $0$ & $-$ & 0 \\\hline
(00,11)$^\dagger$ & 0 & $0$ & $-$ & 1/16 \\\hline
(00,00) & 0 & $-8$ & $2^6$ & 289/64 \\\hline
(SS,SS) & 0 & $-8$ & $2^{10}$ & $(17-8x^2)^2/4$ \\\hline
(S0,S0) & 0 & $-8$ & $2^9$ & $(13+2x^2)^2/9$ \\\hline
(SS,00) & 0 & $0$ & $-$ & $(1-4x^2)^2/72$ \\\hline
(SS,S0) & 0 & $0$ & $-$ & $-8x^4$ \\\hline
\end{tabular}
\caption{Some properties of the hepidd cross sections. The second column gives the power of $E$ \textit{before} the mass cancellations reduce this power to $0$. A factor of $G^2m_G^4$ is implicit in the last two columns and $x=m_S/m_G$. Residues are only given for the leading $\sin(\theta)^{-8}$ behavior. $^\dagger$Processes involving two $S$'s rather than two $0$'s can be obtained from these rows by multiplying values in the last two columns by four.}\label{t1}
\end{table}

In Table \ref{t1} we display some of the properties of the hepidd cross sections, including their growth with $E$ before mass cancellations. All the processes that show the $\sin(\theta)^{-8}$ behavior include elastic scattering, that is where the states remain the same after scattering. The $\sin(\theta)^{-8}$ corresponds to $1/(t^4 u^4)$. Certainly a $\sin(\theta)^{-4}$ behavior is expected from graviton exchange as seen in the graviton scattering result in (\ref{e3}). For finite $E$, the remaining $\sin(\theta)^{-4}$ factor may be regulated by the ghost mass for $\sin(\theta)\lesssim m_G/E$. This is seen to occur in examples where we can inspect the full result, such as in the $11\to 11$ or $00\to 00$ amplitudes. Nevertheless at large $E$ the approximate $\sin(\theta)^{-8}$ behavior means that the elastic scattering completely dominates for near forward or backward scattering. And so it is interesting that we find that residues of the $\theta^{-8}$ and $(\pi-\theta)^{-8}$ poles are always positive, and, strikingly, always just given by a power of two.

Other than $(SS,S0)$, all processes not involving spin-2 have an even number of ghosts, and thus their differential cross sections are intrinsically positive. The ones involving spin-2 have a sum over even and odd numbers of ghosts, and so for them positivity is not guaranteed. It is thus of interest to consider the values of the hepidd cross sections in the transverse direction at $\theta\sim\pi/2$. Here the positive values dominate in Table \ref{t1}, and there are only three negative values. The completely inclusive differential cross section will be positive definite, with a minimum at $\theta=\pi/2$ around which it is symmetric.

In Appendix \ref{a3} we give some examples of differential cross sections before expanding in $1/E$ and thus before mass cancellations are considered. In Appendix \ref{a4} we show how the leading mass cancellation works, which reduces an $E^4$ behavior of hepidd cross sections to $E^2$ behavior.

\section{Some simple amplitudes}\label{s7}

The two-body decay amplitudes among graviparticles are simple. The kinematically allowed decays are
\begin{align}
&G\to gg,\quad S\to gg,\\
&S\to GG\quad\mbox{for }m_S>2m_G,\\
&G\to SS\quad\mbox{for }m_G>2m_S,
\end{align}
where the ghost $G$ has 2, 1 or 0-spin states. The tree-level on-shell amplitudes for all these decays vanish except for one, $0\to SS$, with an amplitude $\propto p_f^2$. For comparison the only decays of graviparticles into real scalars are $0\to \phi\phi$ with amplitude $\propto p_f^2/2$ and $S\to \phi\phi$ with amplitude $\propto (p_f^2+\frac{3}{2}m^2)/\sqrt{2}$. When off-shell decay to on-shell particles is considered then the following processes open up,
\begin{align}
&0\to gg,\quad 0\to 0 0,\\
&0\to 2 2,\quad 2\to 0 2,\\ 
&0\to 1 1,\quad 1\to 0 1,\\ 
&2\to 11,\quad 1\to 21.
\end{align}
These include all placements of $e$ and $o$ with an even number of $o$'s, and there are additional processes with any 0 replaced by an $S$. All these amplitudes are nonvanishing at the respective threshold ($p_f=0$) except for $0\to gg$, $S\to gg$, and $0\to SS$ that are $\propto p_f^2$. Off-shell cubic amplitudes are relevant for a parton-shower process.

Returning to 2-to-2 scattering, we can consider crossing symmetry. A crossing corresponds to changing two momenta so as to interchange two Mandelstam variables, as well as interchanging the two polarization tensors. We have noted that the hepidd cross sections listed in Tables \ref{t2} and \ref{t3} have a $t\leftrightarrow u$ symmetry. But when one particle is in the initial state and the other is in the final state then the two polarization tensors are distinctly different, so interchanging the two Mandelstam variables and not the polarization tensors may not properly implement the crossing. Thus even when the four particles are the same, we might not find a symmetry under interchange of any of $(s,t,u)$. On the other hand when the four particles are identical scalars, we may still expect to find the complete $(s,t,u)$ interchange symmetry.

In the Table \ref{t3} the $(00,00)$ and $(SS,SS)$ entries each involve the square of a single amplitude.\footnote{The scattering amplitudes given in this section all omit a common factor of $8\pi Gm_G^2$.} These amplitudes for $00\to 00$ and $SS\to SS$ may be expressed in terms of Mandelstam variables as follows with $x=m_S/m_G$,
\begin{align}
00\to 00:\quad &\frac{s^{4}}{4 t^{2} u^{2}}+\frac{u^{4}}{4 t^{2} s^{2}}+\frac{t^{4}}{4 s^{2} u^{2}}-2,\label{e14}\\
SS\to SS:\quad &\frac{s^{4}}{t^{2} u^{2}}+\frac{u^{4}}{t^{2} s^{2}}+\frac{t^{4}}{s^{2} u^{2}}-4 x^{2}-8.\label{e15}
\end{align}
We see that they have the complete $(s,t,u)$ symmetry. These high-energy amplitudes have a pole structure that is unlike the scattering amplitudes from any other known QFT.

Crossing also relates $SS\to 0S$ to $0S\to SS$, as well as $S\phi\to 0\phi$ to $0\phi\to S\phi$, but this is trivial because these amplitudes are constants ($y=m/m_G$),
\begin{align}
SS\to 0S:&\quad \sqrt{2}x^2,\\
S\phi\to 0\phi:&\quad \sqrt{2}(2y^2+x^2)/12.
\end{align}

Two of the amplitudes for $11\to 11$ involve four identical particles (four $1e$'s or four $1o$'s) but these amplitudes don't have the complete $(s,t,u)$ symmetry,
\begin{align}
(1e)(1e)\to (1e)(1e):\quad &\frac{s^{4}}{4 t^{2} u^{2}}+\frac{u^{4}}{4 t^{2} s^{2}}+\frac{t^{4}}{4 s^{2} u^{2}}+\frac{(t-u)^2}{s^2}-1/2.
\end{align}
The symmetry is spoiled by the $(t-u)^2/s^2=\cos(\theta)^2$ term that only has the $t\leftrightarrow u$ symmetry. The hepidd cross section for $11\to 11$ sums the squares of amplitudes with all placements of $e$ and $o$. The $(11,11)$ entry in Table \ref{t3} can be written to put most of the pole structure into symmetric terms,
\begin{align}
&\frac{1}{2}\left(\frac{s^{4}}{t^{2} u^{2}}-3\right)^{2}+\frac{1}{2}\left(\frac{u^{4}}{t^{2} s^{2}}-3\right)^{2}+\frac{1}{2}\left(\frac{t^{4}}{s^{2} u^{2}}-3\right)^{2}\nonumber\\
&+\frac{48 t^{6}+121 u \,t^{5}+388 u^{2} t^{4}+470 u^{3} t^{3}+388 u^{4} t^{2}+121 u^{5} t +48 u^{6}}{2  s^{4}u t}.
\end{align}
For the hepidd cross sections that involve mass cancellations, that is that involve spin-2, we do not find any similar patterns.

The $0\phi\to 0\phi$ and $S\phi\to S\phi$ amplitudes do not have $t\leftrightarrow u$ symmetry and they can be expressed as
\begin{align}
0\phi\to 0\phi:&\quad-\frac{\left(y^{2}+\frac{1}{2}\right) s^{2}+\left(2 y^{2}-2\right) s u +\left(y^{2}+\frac{1}{2}\right)u^{2}}{6t^2},\\
S\phi\to S\phi:&\quad-\frac{\left(x^2+y^{2}-\frac{1}{2}\right) s^{2}+\left(2x^2+2 y^{2}+2\right) s u +\left(x^2+y^{2}-\frac{1}{2}\right)u^{2}}{3t^{2}}.
\end{align}
Here the graviparticle pole structure is only in the $t$-channel and we see the expected $s\leftrightarrow u$ symmetry. The $1\phi\to 1\phi$ amplitudes (two $1e$'s or two $1o$'s) also happen to have this symmetry,
\begin{align}
(1e)\phi\to (1e)\phi:&\quad-\frac{s^2+s u+u^2}{2t^2}.
\end{align}

Amplitudes from QQG for $A\phi\to B\phi$ processes were reported in \cite{Abe:2020ikj}, with a focus on how the optical theorem is satisfied. Among the terms in the optical theorem, they observe what we would call the leading mass cancellation, but not the next-to-leading mass cancellations that we also observe. They also observe what we are calling the intrinsic cancellations, but their amplitudes involving 0 or $S$ differ from ours. Our results receive support from the relations we have found between the $A\phi\to B\phi$ processes and the four graviparticle processes.

Modern on-shell methods for amplitudes have made great progress at understanding the whole space of consistent amplitudes in four dimensions \cite{Cheung:2017pzi,Arkani-Hamed:2017jhn}. But it seems that we are well outside this space, given the double poles appearing in our amplitudes, and the fact that we are discussing a massive spin-2 particle. How can this be? First of all we are dealing with an intrinsically massive theory and we are showing amplitudes in the high-energy limit. It is then that the $1/q^2-1/(q^2-m_G^2)$ structure collapses to a double pole. More to the point, perturbative QQG lacks positivity, which is why we are able to have the $1/q^2-1/(q^2-m_G^2)$ structure and a massive spin-2 particle in the first place. The on-shell methods build up amplitudes from the pole contributions in each channel and a ghostlike sign for a pole is presumably never considered. Unitarity is a key requirement for these methods, but as we have argued, unitarity and positivity are distinct properties and it is only positivity that is lost in QQG. So it remains to be seen what happens with the on-shell methods if positivity is dropped. In fact the astonishing cancellations we are finding would seem to indicate that there exists a more direct approach.

\section{Gravitational parton showers}\label{s4}

Scattering at high energies in an asymptotically-free gauge theory, far above the scale of any mass involved, will inevitably involve parton showers. Cross sections for processes that involve extra nearly collinear or soft emissions are enhanced by large logs. While cancellations between the real and virtual parts control IR divergences, the result ends up favoring higher-order diagrams with many branchings. This implies the occurrence of both initial-state and final-state parton showers, as is well observed in the case of high-energy proton-proton scattering.

The inevitability of parton showers in gravity requires some further comment. When a graviton is attached to an external massless particle, the effect is described by a new propagator and a new cubic amplitude. When the new propagator approaches on-shell while simultaneously the two final-state particles approach the collinear limit, the associated cubic amplitude approaches an on-shell amplitude. This on-shell amplitude happens to vanish, both in GR and in QQG, unlike the case when a gauge field is attached. This might lead one to conclude that gravitational parton showers do not occur. But when a massive graviparticle is attached, the corresponding cubic amplitude in the collinear limit is not an on-shell amplitude, and it does not vanish.

The intermediate partons in a parton shower are typically far off-shell, and we saw in the previous section that an off-shell graviparticle can readily split into two graviparticles. The authors of \cite{Salvio:2018kwh} have also argued that there are both soft and collinear singularities in QQG, based largely on the infrared behavior of the quartic propagator. Thus we conclude that gravitational parton showers are inevitable for ultra-Planckian scattering. These parton showers will also involve standard model particles. 

The collision of two particles of any type with ultra-Planckian energies will involve two initial-state gravitational parton showers. These showers evolve through repeated parton splittings, with the new partons having less energy and more spacelike virtuality for every subsequent splitting. These initial-state parton showers interact with each other and produce multiparton interactions. Essentially, the virtual self-energy cloud of one particle is interacting with the virtual self-energy cloud of the other particle. When at least one parton interaction emits sufficient energy in the transverse directions then it can be said that a hard scattering has occurred. For any such hard process the final-state parton showers begin. These showers start with the initial partons having high timelike virtuality. These showers evolve with parton splittings producing partons of ever decreasing energy and virtuality. Eventually the virtuality of the gravitational parton shower has been reduced to the Planck scale. On these time scales the propagation of any ghost is responding to its self-energy corrections, causing it to decay backward in time. This is the physics particular to QQG that replaces the hadronization of QCD.

It is instructive to see how perturbative QCD is used to treat the example of inclusive production of hadrons from high-energy $e^+e^-$ scattering, via the production of QCD jets. The calculation proceeds by finding the cross section for the production of on-shell partons. This is seemingly quite different from the actual process where the partons are initially produced with high timelike virtuality. But it is argued that the probability for the conversion of the inclusive on-shell partonic state to any possible hadronic state is unity and so
\begin{align}
\sigma(e^+e^-\to\textrm{ QCD jets }\to\textrm{ hadrons})=\sigma(e^+e^-\to \textrm{ on-shell QCD partons})
.\end{align}
The inclusive cross section on the rhs is calculated in perturbation theory, with the only requirement being that $E$ be sufficiently above the QCD scale.

By choosing leptons in the initial-state, this example has side stepped the complication of initial-state parton showers. But this is not possible in the gravity case. As with proton-proton scattering, the partonic cross sections must be convoluted with the parton distribution functions of the initial-state showers to obtain physical cross sections. The important point for us is that the partonic cross sections of interest are inclusive with respect to both the initial and final states of the hard scattering process. At a more diagrammatic level, it is off-shell propagators that are linking the hard partonic process with the rest of the process. For a graviparticle it is the propagator in (\ref{e11}) that describes the propagation of all the different degrees of freedom. This translates into the inclusive nature of the hard process, and so it is inclusive differential cross sections that are needed to describe high-energy scattering. Fortunately QQG is able to provide sensible versions of these.

We have described how scattering at large $s$ leads to off-shell processes characterized by high virtuality $|q^2|$, meaning that the off-shell propagators are suppressed due to the near cancellation in $1/q^2-1/(q^2-m_G^2)$. The resulting $-m_G^2/(q^2)^2$ behavior goes along with the higher powers of momenta that appear in the vertices involving graviparticles, and it is these competing effects that control the high-energy behavior. So if we are to use an on-shell picture then it must capture some analogous suppression to balance the vertex behavior. We are finding this suppression at the level of adding exclusive differential cross sections at high $s$. This is what we are referring to as mass cancellations. The root cause of these cancellations is the same in the two pictures; a negative exclusive differential cross section is due to the negative norm of an external ghost, while in the off-shell description the same ghost is represented by the negative ghost propagator. Thus the signs of exclusive cross sections could be viewed as a book-keeping device that keeps track, in the on-shell description, of the same minus signs that occur at the amplitude level in the off-shell description. In both descriptions the cancellations have the effect of maintaining good high-energy behavior. Thus rather than being problematic, the existence of negative exclusive differential cross sections is a crucial and necessary part of the on-shell description.

\section{Conclusion}\label{s8}

In ultra-Planckian scattering, all the elementary gravitational degrees of freedom can be excited whether or not they correspond to true asymptotic states of the theory. We have thus calculated differential cross sections to describe hard scatterings among graviparticles. For large angle scattering, that is small impact parameter, the energy or momentum of the exchanged graviparticle is well above its mass and we are in the asymptotically-free regime. As $\sin(\theta)$ decreases we have found very strong forward and backward enhancement, especially in the case of elastic scattering. When $\sin(\theta)$ becomes smaller than $m_G/E$ then the momentum exchange is smaller than the mass and the differential cross sections become dominated by massless graviton exchange in the $t$ or $u$ channels. Then we enter a regime of GR, which is well known to apply to ultra-Planckian scattering as long as the impact parameter is large enough. 

We have argued that the differential cross sections for hard scattering take their proper place within the context of gravitational parton showers, where the showers occur in both the initial and final states. This is a description involving the off-shell propagation of gravitational degrees of freedom, and it suffers from no conceptual problems in QQG. It is only when one makes the seemingly less physical but more useful on-shell calculations that negative norms enter the picture. The resulting minus signs that occur at the exclusive cross section level do nothing more than mimic the minus signs occurring among propagators at the amplitude level in the off-shell description. In both descriptions, cancellations occur that ensure a well-behaved high-energy behavior.

Thus we are proposing a perturbative description of arbitrarily high-energy scattering. High-energy particles propagate and interact as in the other QFT descriptions of nature. The Planck scale is offering no barrier to high energies, no more than the QCD scale does. The Planck length is simply a distance where the effective low energy theory, i.e.~GR, is breaking down. It is not acting as a minimum length. A whole new perturbative regime opens up at smaller distances, and these distances can be probed by large angle scattering at arbitrarily high energies. This takes place on a spacetime continuum having a normal metric description. This basic picture is protected by a weak, asymptotically-free coupling.

The standard view is that ultra-Planckian scattering will produce black holes, at least for small enough impact parameter. For a final state describable within GR this may be the only possibility, but an asymptotically-free UV-complete QFT offers a different possibility. The QFT does so by offering an explicit microscopic picture for the scattering. An explicit microscopic picture for black hole production is lacking, and the sudden appearance of enormous entropy is mysterious (but see \cite{Dvali:2014ila}). In fact black holes themselves may become less credible in the context of this UV completion. Turning to QCD again, that theory supports (depending on the strange quark mass) macroscopically large solutions in the form of quark matter states, such as quark matter stars. These are ground states of QCD and they cannot be deduced within the domain of the low energy theory, the chiral Lagrangian. Similarly, the true macroscopically large solutions of gravity that act as the endpoint of gravitational collapse might not be properly understood within the realm of GR. Indeed, the study of static solutions of the classical approximation to QQG does indicate the existence of a horizonless replacement to the black hole that is externally very similar \cite{Holdom:2002xy,Holdom:2016nek,Holdom:2019ouz,Ren:2019afg}. The interior solution shows that QQG responds to sufficiently dense matter by reducing the spacetime volume element $\sqrt{-g}$ more dramatically than in GR, in a way that does not yield a horizon. (Inside quark matter, it is the chiral condensate that is dramatically reduced, which is again something not described by the low-energy theory.)

Thus while a description of ultra-Planckian scattering may not be of direct experimental interest, the existence of a UV-complete quantum field theory for gravity offers a change in perspective that is interesting to contemplate.

\appendix
\section{SETUP}\label{a1}

For a graviparticle with mass $m$ and momentum $k_\mu$ we have the following orthogonal 4-vectors,
\begin{align}
k_\mu&=\left(\sqrt{k^2+m^2},k\cos(\theta),k\sin(\theta),0\right)\\
l_\mu&=\left(\frac{k}{m},\frac{\sqrt{k^2+m^2}}{m}\cos(\theta),\frac{\sqrt{k^2+m^2}}{m}\sin(\theta),0\right)\\
t_\mu&=\left(0,-\sin(\theta),\cos(\theta),0\right)\\
u_\mu&=\left(0,0,0,1\right)
\end{align}
Then the polarization tensors can be constructed as follows \cite{Abe:2020ikj}:
\begin{align}
e_{\mu\nu}^{(2,e)}&=\frac{1}{\sqrt{2}}\left(t_\mu t_\nu-u_\mu u_\nu\right)\quad e_{\mu\nu}^{(2,o)}=\frac{1}{\sqrt{2}}\left(t_\mu u_\nu+u_\mu t_\nu\right)\\
e_{\mu\nu}^{(1,e)}&=\frac{1}{\sqrt{2}}\left(l_\mu t_\nu+t_\mu l_\nu\right)\quad e_{\mu\nu}^{(1,o)}=\frac{1}{\sqrt{2}}\left(l_\mu u_\nu+u_\mu l_\nu\right)\\
e_{\mu\nu}^{(0)}&=\frac{1}{\sqrt{6}}\left(2l_\mu l_\nu-t_\mu t_\nu-u_\mu u_\nu\right)\\
e_{\mu\nu}^{(S)}&=\frac{1}{\sqrt{3}}\left(l_\mu l_\nu+t_\mu t_\nu+u_\mu u_\nu\right)=-\frac{1}{\sqrt{3}}\left(\eta_{\mu\nu}-\frac{k_\mu k_\nu}{m^2}\right)
\end{align}
The number of $o$ labels in any amplitude must be even.

The cubic and quartic graviparticle vertices must be obtained under conditions more general than in GR; namely for the metric perturbation $h_{\mu\nu}$ we cannot assume $\partial^2 h_{\mu\nu}= 0$ or $h^\mu_\mu= 0$. We use the Mathematica package xAct`xPerm` \cite{Brizuela:2008ra} to derive the vertices, but they are far too lengthy to display.

\section{TWO-$\phi$-TWO-GRAVIPARTICLE PROCESSES} \label{a2}
Let us first consider the scattering of two scalars $\phi$ to two graviparticles. Two of the diagrams have a scalar exchange in the $t$ and $u$ channels respectively, one has a graviparticle in the $s$ channel along with the cubic graviparticle vertex, and the fourth has the contact interaction. If we let $\cal G$ denote $(g,G)$ and $\cal G^S$ denote $(g,G,S)$, then for the process $\phi\phi\to {\cal GG}$ we find
\begin{align}
\left.\frac{d\sigma_{\phi\phi\to {\cal GG}}}{d\Omega}\right|_{E\to\infty}&=\frac{G^2}{4E^2}\left[I_1+I_2\right],\label{e8}\\\nonumber
I_1&=\frac{m_{2}^{4} \left(-7 \cos \! \left(\theta \right)^{4}+22\cos \! \left(\theta \right)^{2}+1\right)}{8},\\\nonumber
I_2&=\frac{\left(3 \cos \! \left(\theta \right)^{2} m_{2}^{2}+4 m^{2}-m_{2}^{2}\right)^{2}}{144}.
\end{align}
This is a result at leading order in $E$ after realizing the mass cancellations. If we instead consider $\phi\phi\to{\cal G}^S{\cal G}^S$ then $I_2$ is replaced by $I_3$,
\begin{align}
I_3&=\frac{\left(-3 \cos \! \left(\theta \right)^{2} m_{2}^{2}+4 m^{2}+m_{2}^{2}+4 m_{0}^{2}\right) \left(-15 \cos \! \left(\theta \right)^{2} m_{2}^{2}+4 m^{2}+5 m_{2}^{2}+12 m_{0}^{2}\right)}{144}.
\end{align}
In terms of the spin content of the two graviparticles, $I_1$ comes from 22, 21, 11, $I_2$ from 00 and $I_3$ from 00, SS, S0. The corresponding result for ${\cal GG}\to\phi\phi$ has $I_1$ replaced by $m_{2}^{4} \left(-9 \cos \! \left(\theta \right)^{4}+10 \cos \! \left(\theta \right)^{2}+15\right)/8$.

We next consider the process ${\cal G}\phi\to{\cal G}\phi$. We find
\begin{align}
\left.\frac{d\sigma^{\rm na}_{{\cal G}\phi\to{\cal G}\phi}}{d\Omega}\right|_{E\to\infty}&=\frac{G^2}{4E^2}\left[J_1+J_2\right],\label{e9}\\\nonumber
J_1&=\frac{8 m_{2}^{4} \left(-\cos \! \left(\theta \right)^{3}+9 \cos \! \left(\theta \right)^{2}+5\cos \! \left(\theta \right)-1\right)}{\left(\cos \! \left(\theta \right)-1\right)^{4}},\\\nonumber
J_2&=\left(\frac{m^{2}}{3}+\frac{m_{2}^{2}}{6}+\frac{2 m_{2}^{2} \left(\cos \! \left(\theta \right)+1\right)}{\left(\cos \! \left(\theta \right)-1\right)^{2}}\right)^{2}.
\end{align}
For ${\cal G}^S\phi\to{\cal G}^S\phi$ then $J_2$ is replaced by $J_3$,
\begin{align}
J_3&=\frac{1}{36}\left(2 m^{2}-5 m_{2}^{2}+6 m_{0}^{2}-\frac{60 m_{2}^{2} \left(\cos \left(\theta \right)+1\right)}{\left(\cos \left(\theta \right)-1\right)^{2}}\right) \left(2 m^{2}-m_{2}^{2}+2 m_{0}^{2}-\frac{12 m_{2}^{2} \left(\cos \left(\theta \right)+1\right)}{\left(\cos \left(\theta \right)-1\right)^{2}}\right).
\end{align}
The contributions in terms of the spin content of the graviparticles is the same as before. Each of these $J_i$'s has a $\theta^{-8}$ pole, and in each case the residue of this pole is positive. In terms of the Mandelstam variable this is $1/t^4$ behavior, to be compared with the $1/t^2$ behavior of the corresponding GR result \cite{Berends:1974gk}.

\section{SOME FULL RESULTS}\label{a3}
As a point of reference we calculate the differential cross section from QQG for the scattering among massless gravitons $gg\to gg$, 
\begin{align}
\frac{d\sigma^{\rm na}_{gg\to gg}}{d\Omega}= G^2 E^2 \frac{(\cos(\theta)^2+3) (\cos(\theta)^6+25 \cos(\theta)^4-5 \cos(\theta)^2+43)}{\sin(\theta)^4}
.\label{e3}\end{align}
The ``na'' indicates that this is not yet averaged over initial polarizations. Doing so introduces an additional factor of 1/4, and then this result is equivalent to the GR result \cite{Berends:1974gk}. The $1/\sin(\theta)^4$ corresponds to $1/(tu)^2$ behavior.

Next we give the full result for two gravitons scattering into two ghosts $gg\to GG$,
\begin{multline}
\frac{d\sigma^{\rm na}_{gg\to GG}}{d\Omega}=\frac{G^2\sqrt{1-m_G^2/E^2}}{2E^2\left(\sin(\theta)^{2}(E^2-m_G^2)+m_G^2 \right)^{2}}
\left[ 5\,(E^2-m_G^2)^4\sin(\theta)^{8}\right.\\
\mbox{}+(-80 E^8+288 E^6 m_G^2-336 E^4 m_G^4+64 E^2 m_G^6)\sin(\theta)^{6}\\
\mbox{}+(336 E^8-864 E^6 m_G^2+400 E^4 m_G^4)\sin(\theta)^{4}\\
\mbox{}+(-512 E^8+640 E^6 m_G^2)\sin(\theta)^{2}\\
\left.\mbox{}+512 E^8-512 E^6 m_G^2+336 E^4 m_G^4-80 E^2 m_G^6+5 m_G^8 \right]
\label{e4}\end{multline}
Here we see no singular $\sin(\theta)^{-4}$ behavior, and so the graviton pole in the $t$ and $u$ channels has been replaced by a ghost pole. This is due to the vanishing of the on-shell $Ggg$ vertex. Some of the amplitudes depend on the graviscalar mass $m_S$, but this dependence cancels out in the final result. And finally we give the scattering into two graviscalars $gg\to SS$,
\begin{align}
\frac{d\sigma^{\rm na}_{gg\to SS}}{d\Omega}=\frac{2G^2\sqrt{1-m_S^2/E^2}}{E^2\left(\sin(\theta)^2 (E^2-m_S^2)+m_S^2\right)^2}\; \left[(E^2-m_S^2)^4 \sin(\theta)^8+m_S^8\right]
.\end{align}

\section{THE LEADING CANCELLATIONS}\label{a4}
In this section we show how the mass cancellation works at leading order in $1/E$, reducing the $E^2$ behavior of dimensionful-differential cross sections to $E^0$ behavior. The less trivial next-to-leading order cancellations that produce the desired $1/E^2$ behavior are not discussed here.

In the high-energy limit the results in the previous section reduce to
\begin{align}
\left.\frac{d\sigma^{\rm na}_{gg\to GG}}{d\Omega}\right|_{E\to\infty}&=\frac{d\sigma^{\rm na}_{gg\to gg}}{d\Omega}+\frac{1}{2} G^2 E^2 (\cos(\theta)^2+3) (3 \cos(\theta)^2+1)\label{e2}\\
\left.\frac{d\sigma^{\rm na}_{gg\to SS}}{d\Omega}\right|_{E\to\infty}&=2G^2E^2\sin(\theta)^4
\end{align}
The first term in (\ref{e2}) is the result in (\ref{e3}), and it arises here when the ghost polarizations are both spin-2. The second term in (\ref{e2}) is due to when the ghost polarizations are either both spin-0 or both spin-1. The $Gg$ final state might have also provided an ${\cal O}(E^2)$ contribution, but that amplitude identically vanishes. This is an example of a nontrivial property of the on-shell amplitudes. Other amplitudes contribute at most at order ${\cal O}(E^0)$.

Thus the high-energy contribution to the scattering of massless gravitons to anything comes from $gg\to gg+GG+SS$ and is
\begin{align}
&\left.\frac{d\sigma^{\rm na}_{gg\to gg+GG+SS}}{d\Omega}\right|_{E\to\infty}={\cal S}\label{e5}\\
&{\cal S}=2\frac{d\sigma^{\rm na}_{gg\to gg}}{d\Omega}+G^2 E^2 \left[\frac{1}{2}(\cos(\theta)^2+3) (3 \cos(\theta)^2+1)+2\sin(\theta)^4\right]
\end{align}
We define $\cal S$ since it will also appear in results below.

Next we consider the scattering of two ghosts. The process $GG\to GG$ gives
\begin{multline}
\left.\frac{d\sigma^{\rm na}_{GG\to GG}}{d\Omega}\right|_{E\to\infty}=9\frac{d\sigma^{\rm na}_{gg\to gg}}{d\Omega}+G^2 E^2 (\cos(\theta)^2+3) (3 \cos(\theta)^2+1)\\\mbox{}+8G^2 E^2\frac{(\cos(\theta)^2+3)(\cos(\theta)^4+16\cos(\theta)^2+7)}{\sin(\theta)^4}
.\end{multline}
To contribute to the leading $E^2$ behavior, at least two of the ghosts must have spin-2 polarization, and the first term is due to all having spin-2 polarization. The factor of nine comes from a factor of three in the high-energy $GG\to GG$ amplitudes as compared to the corresponding amplitudes in $gg\to gg$. This is another example of nontrivial on-shell behavior. The second term is due to when either the initial or final state has two spin-1 or two spin-0 polarizations, while the third is due to when the initial and final states each have one spin-1 or each have one spin-0. We notice a common factor of $\cos(\theta)^2+3$ in all three terms.

The ghosts may also scatter into two massless gravitons or two graviscalars where
\begin{align}
\left.\frac{d\sigma^{\rm na}_{GG\to gg}}{d\Omega}\right|_{E\to\infty}&=\left.\frac{d\sigma^{\rm na}_{gg\to GG}}{d\Omega}\right|_{E\to\infty},\\
\left.\frac{d\sigma^{\rm na}_{GG\to SS}}{d\Omega}\right|_{E\to\infty}&=\left.\frac{d\sigma^{\rm na}_{gg\to SS}}{d\Omega}\right|_{E\to\infty}
.\end{align}
A negative cross section appears when we consider $GG\to Gg+gG$. We find
\begin{multline}
\left.\frac{d\sigma^{\rm na}_{GG\to Gg+gG}}{d\Omega}\right|_{E\to\infty}=-8\frac{d\sigma^{\rm na}_{gg\to gg}}{d\Omega}- G^2 E^2 (\cos(\theta)^2+3) (3 \cos(\theta)^2+1)\\\mbox{}-8G^2 E^2\frac{(\cos(\theta)^2+3)(\cos(\theta)^4+16\cos(\theta)^2+7)}{\sin(\theta)^4}
.\end{multline}
The factor of 8 in the first term comes from a factor of two in the amplitudes and a factor of two from the final state. Thus the high-energy contribution of the scattering of ghosts into anything comes from $GG\to gg+GG+Gg+gG+SS$ and it ends up as
\begin{align}
\left.\frac{d\sigma^{\rm na}_{GG\to gg+GG+Gg+gG+SS}}{d\Omega}\right|_{E\to\infty}={\cal S}
\label{e6}.\end{align}

Next we may consider the scattering of a ghost with a massless graviton, where we find
\begin{align}
\left.\frac{d\sigma^{\rm na}_{Gg\to Gg+gG}}{d\Omega}\right|_{E\to\infty}=2\frac{d\sigma^{\rm na}_{gg\to gg}}{d\Omega}+4G^2 E^2\frac{(\cos(\theta)^2+3)(\cos(\theta)^4+16\cos(\theta)^2+7)}{\sin(\theta)^4}
.\end{align}
We also have
\begin{align}
\left.\frac{d\sigma^{\rm na}_{Gg\to GG}}{d\Omega}\right|_{E\to\infty}=\left.\frac{1}{2}\frac{d\sigma^{\rm na}_{GG\to Gg+gG}}{d\Omega}\right|_{E\to\infty}
.\end{align}
The amplitude for $Gg\to gg$ vanishes, and thus the high-energy contribution to the scattering of $Gg$ into anything comes from $Gg\to GG+Gg+gG+SS$ and is
\begin{align}
\left.\frac{d\sigma^{\rm na}_{Gg\to GG+Gg+gG+SS}}{d\Omega}\right|_{E\to\infty}=-{\cal S}
\label{e7}.\end{align}

We have found in (\ref{e5}), (\ref{e6}) and (\ref{e7}) that the total unaveraged cross sections in the high-energy limit for four different initial states, $gg$, $GG$, $Gg$, and $gG$ are all the same, up to a sign. This is a nontrivial result, since the various contributions and the sets of polarizations that are involved in each case is different. If we let $\cal G$ denote $(g,G)$ and $\cal G^S$ denote $(g,G,S)$, then we can consider the inclusive process ${\cal GG}\to {\cal G}^S{\cal G}^S$,
\begin{align}
\left.\frac{d\sigma^{\rm na}_{{\cal GG}\to {\cal G}^S{\cal G}^S}}{d\Omega}\right|_{E\to\infty}={\cal S+S-S-S} +{\cal O}
(E^0)={\cal O}(E^0)
.\end{align}
We see that the leading ${\cal O}(E^2)$ terms cancel. In the main text we point out that the next-to-leading ${\cal O}(E^0)$ terms also cancel, leaving $1/E^2$ behavior. At next-to-leading order the different angular dependences are more complicated and it is best to show the cancellations for each minimally-inclusive process separately.

\acknowledgments
I thank John Donoghue and Richard Woodard for an interesting email exchange concerning various big picture views. I acknowledge the Natural Sciences and Engineering Research Council of Canada (NSERC) for partial support.


\begin{thebibliography}{99}

\bibitem{stelle}
 K.~S.~Stelle,    Renormalization   of   higher   derivative   quantum   gravity,    Phys.~Rev.~D \textbf{16},    953   (1977).

\bibitem{julve}
 J.~Julve   and   M.~Tonin,   Quantum   gravity   with   higher   derivative   terms,   Nuovo   Cim.~B \textbf{46},   137   (1978).
 
\bibitem{fradkin}
 E.~S.~Fradkin and A.~A.~Tseytlin, Renormalizable asymptotically free quantum theory of gravity, Phys.~Lett.~ \textbf{104}B, 377 (1981); E.~S.~Fradkin and A.~A.~Tseytlin, Renormalizable asymptotically-free quantum theory of gravity, Nucl.~Phys.~B\textbf{201}, 469 (1982).
 
\bibitem{Donoghue:2019ecz}
J.~F.~Donoghue and G.~Menezes,
Arrow of causality and quantum gravity,
Phys. Rev. Lett. \textbf{123}, 171601 (2019).

\bibitem{Donoghue:2020mdd}
J.~F.~Donoghue and G.~Menezes,
Quantum causality and the arrows of time and thermodynamics,
Prog. Part. Nucl. Phys. \textbf{115}, 103812 (2020).
 
\bibitem{leewick}
T.~D.~Lee  and  G.~C.~Wick,  Negative  metric  and  the  unitarity  of  the  $S$-matrix,  Nucl.~Phys.~B\textbf{9},  209  (1969); T. D. Lee and G. C. Wick, Unitarity in the $N\theta\theta$ sector of soluble model with indefinite metric, Nucl. Phys. B\textbf{10}, 1(1969);  T.~D.~Lee   and   G.~C.~Wick,   Finite   theory   of   Quantum   Electrodynamics,   Phys.~Rev.~D \textbf{2},   1033   (1970).

\bibitem{Grinstein:2008bg}
B.~Grinstein, D.~O'Connell and M.~B.~Wise,
Causality as an emergent macroscopic phenomenon: the Lee-Wick O(N) model,
Phys. Rev. D \textbf{79}, 105019 (2009).

\bibitem{Donoghue:2019fcb}
J.~F.~Donoghue and G.~Menezes,
Unitarity, stability and loops of unstable ghosts,
Phys. Rev. D  \textbf{100}, 105006 (2019).

\bibitem{Salvio:2018kwh}
A.~Salvio, A.~Strumia and H.~Veerm\"ae,
New infra-red enhancements in 4-derivative gravity,
Eur. Phys. J. C \textbf{78}, 842 (2018).

\bibitem{Bender:2007wu}
C.~M.~Bender and P.~D.~Mannheim,
No-ghost theorem for the fourth-order derivative Pais-Uhlenbeck oscillator model,
Phys. Rev. Lett. \textbf{100}, 110402 (2008).

\bibitem{Salvio:2018crh}
A.~Salvio,
Quadratic gravity,
Front. in Phys. \textbf{6}, 77 (2018).

\bibitem{Strumia:2017dvt}
A.~Strumia,
Interpretation of quantum mechanics with indefinite norm,
MDPI Physics \textbf{1}, 17 (2019).

\bibitem{Salvio:2020axm}
A.~Salvio,
Dimensional transmutation in gravity and cosmology,
Int. J. Mod. Phys. A \textbf{36}, 2130006 (2021).

\bibitem{Holdom:2019ouz} 
  B.~Holdom,
  A ghost and a naked singularity; facing our demons,
  arXiv:1905.08849 [gr-qc].
  
\bibitem{Holdom:2015kbf}
B.~Holdom and J.~Ren,
QCD analogy for quantum gravity,
Phys. Rev. D \textbf{93}, 124030 (2016).

\bibitem{Holdom:2016xfn}
B.~Holdom and J.~Ren,
Quadratic gravity: from weak to strong,
Int. J. Mod. Phys. D \textbf{25}, 1643004 (2016).

\bibitem{Feyn}
R.P.~Feynman, Negative probability in quantum implications: Essays in honor of David Bohm, edited by B.J.~Hiley and F.D.~Peat (Routledge and Kegan Paul, London, 1987), Chap. 13, pp 235-248, https://cds.cern.ch/record/154856/files/pre-27827.pdf

\bibitem{Salvio:2017qkx}
A.~Salvio and A.~Strumia,
Agravity up to infinite energy,
Eur. Phys. J. C \textbf{78}, 124 (2018).

\bibitem{Dona:2015tra}
P.~Don\`a, S.~Giaccari, L.~Modesto, L.~Rachwal and Y.~Zhu,
Scattering amplitudes in super-renormalizable gravity,
JHEP \textbf{08}, 038 (2015).

\bibitem{Berends:1974gk}
F.~A.~Berends and R.~Gastmans,
On the high-energy behavior in quantum gravity,
Nucl. Phys. B\textbf{88}, 99 (1975).

\bibitem{Abe:2020ikj}
Y.~Abe, T.~Inami and K.~Izumi,
Perturbative S-matrix unitarity ($S^\dagger S = 1$) in $R^2_{\mu\nu}$ gravity,
Mod. Phys. Lett. A \textbf{36}, 2150105 (2021).

\bibitem{Cheung:2017pzi}
C.~Cheung,
TASI lectures on scattering amplitudes, Anticipating the Next Discoveries in Particle Physics (World Scientific, Singapore, 2018)
[arXiv:1708.03872 [hep-ph]].

\bibitem{Arkani-Hamed:2017jhn}
N.~Arkani-Hamed, T.~C.~Huang and Y.~t.~Huang,
Scattering amplitudes for all masses and spins,
JHEP \textbf{11}, 070 (2021).

\bibitem{Dvali:2014ila}
G.~Dvali, C.~Gomez, R.~S.~Isermann, D.~L\"ust and S.~Stieberger,
Black hole formation and classicalization in ultra-Planckian $2\to N$ scattering,
Nucl. Phys. B\textbf{893}, 187 (2015).

\bibitem{Holdom:2002xy}
B.~Holdom,
On the fate of singularities and horizons in higher derivative gravity,
Phys. Rev. D \textbf{66}, 084010 (2002).

\bibitem{Holdom:2016nek}
B.~Holdom and J.~Ren,
Not quite a black hole,
Phys. Rev. D \textbf{95}, 084034 (2017).

\bibitem{Ren:2019afg}
J.~Ren,
Anatomy of a thermal black hole mimicker,
Phys. Rev. D \textbf{100}, 124012 (2019).

\bibitem{Brizuela:2008ra}
D.~Brizuela, J.~M.~Martin-Garcia and G.~A.~Mena Marugan,
xPert: Computer algebra for metric perturbation theory,
Gen. Relativ. Gravit. \textbf{41}, 2415 (2009).

\end{thebibliography}
\end{document}